\documentclass[twocolumn,showpacs,preprintnumbers,amsmath,amssymb]{revtex4}
\newcommand{\Sc}{Schr\"odinger }
 \newcommand{\al}{\alpha }
 \newcommand{\vfi}{\varphi }

 \usepackage{epsfig}

\begin{document}

\title{Supersymmetry approach to nuclear-spin-polarization-induced quantum dot structure calculations}

\author{Boris F. Samsonov}
\affiliation{
Departamento de F\'{\i}sica Te\'orica,\\ Universidad de
Valladolid,  47005 Valladolid, Spain and\\
Physics Department of Tomsk State
 University, 36 Lenin ave., 634050 Tomsk, Russia
}

\author{Yuriy V. Pershin}
\affiliation{Center for Quantum Device Technology,\\ Department of
Physics, Clarkson University, Potsdam, NY 13699-5721, USA}

\begin{abstract}
In nuclear-spin-polarization-induced quantum dots the electrons
are confined through local nuclear spin polarization.
The model electron
confinement potential is time-dependent due to the nuclear spin
diffusion and relaxation processes.
It can be well-approximated by a Gaussian curve which is not an
exactly solvable potential.
We demonstrate that it can also be approximated by multisoliton
potentials for the zero value of the angular momentum and by their
singular analogues for other values of momentum without any loss of
calculational accuracy.
We
obtain these potentials by supersymmetric (or equivalently Darboux)
 transformations from the zero potential. The main advantage of such potentials is
that they are exactly solvable.
 Time-dependence of the
nuclear-spin-polarization-induced quantum dot energy levels is
found.

\end{abstract}


\maketitle

\section{\label{intro}Introduction}

In recent times theoretical and experimental investigations of
quantum dots is attracting a considerable attention
\cite{Jacak,Peeters4,McCord,Leadbeater91,Bending,
Peeters1,Solimany,Kim1,Leburton,FIPV02,QD, SOQD}. In particular,
it has been suggested that a novel class of quantum dots
\cite{FIPV02,QD}, so-called nuclear-spin-polarization-induced
(NSPI) quantum dots (QD), as well as other NSPI low-dimensional
electron structures \cite{QW,QR,LMKHV03}, could be created through
locally polarized nuclear spins. The main idea of creating NSPI
structures consists of the following. Electron and nuclear spins
interact via the contact hyperfine interaction that can be
described by an effective hyperfine field ${\bf B}_{hf}$, which
acts at spins of electrons and contributes to the electronic
Hamiltonian through a Zeeman-type potential $g^* \mu _{B}{\bf
\sigma B}_{hf}\left( {{\bf r},t}\right)/2 $ \cite{Slichter}. (Here
$g^*$ is the effective electron $g$-factor, $\mu _{B}$ is the Bohr
magneton and ${\bf \sigma }$ is the Pauli matrix-vector). The
Zeeman splitting results in the potential that is attractive for
the electrons with one spin projection and repulsive for others.
In this model the energy is shifted by a constant of the order of
the Fermi energy by means of a gate. Then the potential, created
by the inhomogeneous nuclear spin polarization
\begin{equation} \label{U_conf}
U_{conf}({\bf r}{,t})=-\frac{\left| g^*\right| }{2}\mu
_{B}B_{hf}\left( {{\bf r },t}\right)
\end{equation}
is confining.
It is assumed that $B_{hf}\left( {{\bf r},t}\right) >0$.

The idea to create NSPI QDs has been proposed for the first time
by Fleurov et al \cite{FIPV02}. They used a perturbation theory to
find the modification of the energy spectrum of a traditional
quantum dot due to the polarized nuclear spins. Formalism
describing NSPI structures was thereafter developed and used in
investigations of NSPI quantum wires \cite{QW}. It was established
that properties of NSPI structures are time-dependent because of
nuclear spin diffusion and relaxation processes. Subsequently,
this formalism has been used to investigate NSPI quantum dots
\cite{QD}, NSPI quantum rings \cite{QR}, and NSPI periodic
structures \cite{LMKHV03}. Moreover, related low-dimensional
electron structures created through modulation of spin-orbit
interaction coupling constant have also been considered \cite{SOQD}.

It was assumed in \cite{QD} that a NSPI QD is created in the
region of the intersection of the two-dimensional electron gas
with cylindrically polarized nuclear spins. The electron energy
levels at any particular time were found as a solution of the 2D
radial Shr\"odinger equation with a Gaussian confining potential.
The 2D radial Shr\"odinger equation (as well as 3D one) with a
Gaussian potential does not give analytic solutions. Different
methods were used to solve this problem
\cite{Gaus1,Gaus2,Gaus3,Gaus4,Gaus5}. A parabolic approximation of
the Gaussian potential has been used in Ref. \cite{QD}. However,
such an approximation is acceptable
 only for the ground state \cite{QD,Gaus1}.

In this paper we extend the study of NSPI systems by considering
NSPI QD created through spherically polarized
 nuclear spins. Time dependence of the electron
confining potential of Gaussian type
 is found as a solution of the diffusion equation
with a relaxation term and the problem is thus reduced to solving
the 3D radial Shr\"odinger equation where the time is involved as
a parameter. For the zero value of the angular momentum we
approximate the Gaussian potential, which is not exactly solvable,
by a multisoliton potential, which is exactly solvable. The
multisoliton potential is obtained by the technique of
supersymmetric quantum mechanics (or equivalently by the method of
Darboux transformations). For other angular momenta the method
gives their singular analogues. It should be mentioned that the
applicability of our approach is not restricted only to NSPI QDs.
It could be used for describing traditional quantum dots
\cite{Gaus1} also.

The paper is organized as follows. We introduce the basic
equations describing NSPI QD in Sec. \ref{basiceq}. In Sec.
\ref{method} we present the method of calculations. Time
dependence of the electron states in NSPI QD is studied in Sec.
\ref{results}. The conclusions are drawn in Sec.
\ref{conclusions}.

\section{\label{basiceq}Basic equations}

Let us consider a semiconductor structure with locally polarized
nuclear spins. There are two main mechanisms leading to the time
dependence of the hyperfine field describing the nuclear spin
polarization: nuclear spin relaxation and nuclear spin
diffusion. Then the evolution of the hyperfine field is governed by the
diffusion equation
\begin{equation}  \label{diffusion}
\frac{\partial B_{hf}}{\partial t}=D\Delta
B_{hf}-\frac{1}{T_{1}}B_{hf}
\end{equation}
accounting also for the relaxation processes \cite{QW}. Here $D$ is the
spin-diffusion coefficient, $\Delta$ is the
usual
three-dimensional Laplacian, and $T_{1}$ is  nuclear
spin relaxation time \cite{Slichter,Wolf}.
In the simplest case,
we can assume
a Gaussian form
$B_{hf}\left( r,0\right) =B_{0}\exp \left(
-\frac{r^{2}}{2d^{2}}\right) $ for
 the initial condition.
 The parameters $d$ and
$B_{0}$ define the half-width and the amplitude of the initial
distribution of the hyperfine field respectively. Then the
solution of Eq. (\ref{diffusion}) is:
\begin{equation}  \label{B(r,t)}
B_{hf}\left( r,t\right) =B_{0}e^{-\frac{t}{T_{1}}}\left(
1+\frac{t}{t_{0}} \right)
^{-\frac{3}{2}}e^{-\frac{r^{2}}{2d^{2}\left( 1+\frac{t}{t_{0}}
\right) }}\,,
\end{equation}
where $t_{0}=\frac{d^{2}}{2D}$. The maximum nuclear field in GaAs
can be as high as $B_{hf}=5.3T$ in the limit that all nuclear
spins are fully polarized \cite{Paget}. This high level of nuclear
spin polarization has been achieved experimentally. For example,
the optical pumping of nuclear spins in
2DEG has demonstrated the
nuclear spin polarization of the order of $90\%$ \cite{Salis}. A
similar high polarization has been created by quantum Hall edge
states ($85\%$) \cite{Dixon}.  These techniques are developed to
allow local polarization and controllability of nuclear spins
\cite{recent}.

Our approach is based on the following
electronic Hamiltonian:
\begin{equation}\label{elham}
H=-\frac{\hbar ^{2}}{2m^{\ast }}\Delta -\frac{\left| g^*\right|
}{2}\mu _{B}B_{hf}\left( {{\bf r },t}\right)\,,
\end{equation}
where $m^{\ast }$ is the electron effective mass. The time scale
introduced by the nuclear spins is several orders of magnitude
larger than the time scale of typical electron equilibration
processes. In such a case the electrons feel a quasi-constant
average nuclear field. This simplifies calculation by avoiding the
complications which would appear when solving the Schr\"{o}dinger
equation with the time dependence due to polarized nuclei. We take
into account only the electrons with the spins along the $z$ axis,
for which the effective potential is attractive.

The electron energy levels in NSPI QD are determined by the radial
Schr\"{o}dinger equation that
in appropriate units takes the form
\begin{equation}\label{Shred}
\left[- \frac{d^2}{dx^2}+V^{(l)}(x,t)- E \right]\psi_{n,l} (x,t) =0\,,
\end{equation}
where $\psi_{n,l} (x,t)$ is related with the radial part of the wave
function $R_{n,l}(x,t)$ as follows: $\psi_{n,l} (x,t)=xR_{n,l}(x,t)$,
\begin{equation}\label{pot}
V^{(l)}(x,t)=\frac{l(l+1)}{x^{2}} -\gamma
\frac{B_{hf}(x,t)}{B_{hf}(0,0)}\,,
\end{equation}
$x=r/d$, $\gamma= \left| g^* \right| \mu_{B}B_{hf}(0,0)/\left(2E_0
\right)$, $E=\varepsilon/E_0$ is the energy in dimensionless
units, $\varepsilon$ is the energy, $E_0=\hbar^2/\left(
2m^*d^2\right)$, and $l=0,1,\ldots$. In the next Section we
describe the method we are using to approximate the potential
(\ref{pot}) with the effective hyperfine field $B$ given by Eq.
(\ref{B(r,t)}).

\bigskip

\section{\label{method}Method of calculation}

The Darboux transformation method also known as the method of
supersymmetric quantum mechanics is
an effective tool for solving different problems of theoretical
and mathematical physics (for reviews see \cite{SUSY}).
Here we are using its property to approximate an interaction
between composite particles by a local potential with an
experimental accuracy, which previously proved to be very
efficient in describing iso-phase
(also known as phase equivalent) potentials in
 nuclear physics \cite{SB,SS}.
 Below we outline shortly the main
 features of the method we need.

Suppose one knows all solutions of the \Sc equation with a given
reference potential $V_0$
\begin{equation}\label{init}
h_0\psi =E\psi ,\quad h_0=-\frac{d^2}{dx^2}+V_0(x)\,.
\end{equation}
Then, using a simple algorithm, one can construct a huge
multiparameter family of exactly solvable potentials. The
solutions $\vfi$ of the \Sc equation with these potentials are obtained by
acting on solutions of $h_0$ with a differential operator,
$\vfi=L\psi$. In the simplest case this is a first order operator
\begin{equation}\label{L1}
L=-d/dx +w(x)\,,
\end{equation}
where the real function $w(x)$
(known in supersymmetric quantum mechanics as {\it superpotential})
 is defined as the
logarithmic derivative of a  solution $u$ of (\ref{init}).
So, one has
\begin{equation}\label{SUPER}
w=u'(x)/u(x)\,,\quad h_0u=\al u\,,
\end{equation}
with $\al \le E_0$, where $E_0$ is the ground state energy of
$h_0$ if it has a discrete spectrum or the lower bound of the
continuous spectrum. The function $u$ is called {\it
transformation} or {\it factorization function} and $\al $ its
{\it factorization constant} or {\it factorization energy}. The
potential $V_1$ of the Hamiltonian $h_1=-d^2/dx^2+V_1$, $h_1\vfi
=E\vfi$, is defined in terms of the superpotential $w$ as
\begin{equation}\label{V1general}
V_1(x)=V_0(x)-2w'(x)\,.
\end{equation}
Eq. (\ref{L1}) defines a first order Darboux transformation. In
the following we shall deal with chains of $N$ successive
transformations of this type.


Here we will use very special chains introduced in \cite{SS}.
They are generated by the following system of
transformation functions
\begin{eqnarray}\label{uv}
v_{1}(x),\ldots , v_l (x),
u_{l +1}(x),  v_{l +1}(x),\ldots , u_n(x), v_n(x)\\
h_0u_j(x)=-a^2_ju_j(x)\,,\quad h_0v_j(x)=-b^2_jv_j(x)\,,\qquad
\end{eqnarray}
where $v_j$ are regular ($v_j(0) = 0$) and
$u_j$ irregular ($u_j(0) \neq 0)$ at the origin.
They have arbitrary eigenvalues $-{a_j}^2$ and  $-{b_j}^2$
respectively, but always below  $E_0$.
If we are interested in the final action of the chain only, the solution
$\psi _N (x,k)$ of the transformed equation with the Hamiltonian
\begin{equation}\label{FINAL}
h_N = -d^2/dx^2+ V_N
\end{equation}
corresponding to the energy $E = k^2$ is given
by \cite{CRUM}
\begin{equation}\label{psiN}
\psi _N (x,k)=
W(u_1,\ldots ,u_N,\psi _0(x,k))\,
W^{-1}(u_1,\ldots ,u_N)
\end{equation}
where $W$ are Wronskians expressed in terms of $u_j$, denoting
symbolically any function of (\ref{uv}) and of $\psi _0(x,k)$
which is a solution of the original Schr\"odinger equation
corresponding to the same energy $E$, $N=2n-l$. In the Hamiltonian
(\ref{FINAL}) the transformed potential is
\begin{equation}\label{VN}
V_N=V_0-2\frac {d^2}{dx^2}\ln W(u_1,\ldots ,u_N)\,.
\end{equation}
For $N=1$ one has $W(u_1)\equiv u_1$
and one recovers (\ref{V1general}) with $u=u_1$.
If $V_0$ is finite at the origin, $V_N$ behaves as
$l (l +1)x^{-2}$ when $x\to 0$.
Therefore the parameter $l$
can be associated with the value of the angular momentum in
(\ref{pot}).
The formulas (\ref{psiN}) and (\ref{VN})
result from the replacement of a chain of $N$ first
order transformations by a single $N$th order transformation,
which happens to be more efficient in practical calculations.

In Ref. \cite{SS} we obtained
that the transformed Jost function $F_N$ is related to the initial
Jost function $F_0$ by
\begin{equation}\label{MNnu}
F_N(k)=
F_0(k)\prod\limits_{j\,=\,1}^{l }\frac{k}{k+ib_j}
\prod\limits_{j\,=\,l+1 }^n\frac{k-ia_j}{k+ib_j}~.
\end{equation}
For $l = 0$ the first product is unity. Since a Jost function is
analytic in the upper half of the complex $k$-plane (see e. g.
\cite{FADDEEV}) all $b$'s must be positive whereas the $a$'s can
have any sign, so that every positive $a_j$ corresponds to a
discrete level $E=-a_j^2$ of $h_N$.

In our case we choose $V_0=0$ so that only exponentials are
involved in the final Hamiltonian. Moreover, since the potential
(\ref{pot}) is symmetric with respect to inversion $x\to -x$ we
take $u_j(x)=\cosh (a_jx)$ and $v_j=\sinh (b_jx)$. In this case
for $l=0$ we get symmetric multisoliton potentials (see e.g. the
first of Refs. \cite{SUSY}). Moreover, the Wronskian from
(\ref{VN}) can be expressed as a sum of hyperbolic cosines
\cite{SamShek}. For $l>0$ we get singular at the origin analogues
of multisoliton potentials. To compare them with the second term
of the right hand side of (\ref{pot}) we have to subtract the
centrifugal part and consider $V_{eff}=V_N-l(l+1)x^{-2}$. The
potential $V_{eff}$ depends on $N=2n-l$ parameters $a_j$ and $b_j$
which we fit to the Gaussian curve of the second term of the right
hand side of (\ref{pot}) for all values of $t$. This gives us the
time dependence of the parameters $a_j$ and $b_j$ and hence the
energy levels $E_j=-a_j^2$. To get an idea how many functions of
type (\ref{uv}) could be involved in a particular case, we plotted
the potentials (\ref{VN}) for $N=2$, $4$, \ldots , $10$, $l=0$,
$B_0=1$, $t_0/T_1=1$, $\tau =0$ in Fig. 1.
\begin{figure}\label{fig1}
\centering \epsfig{file=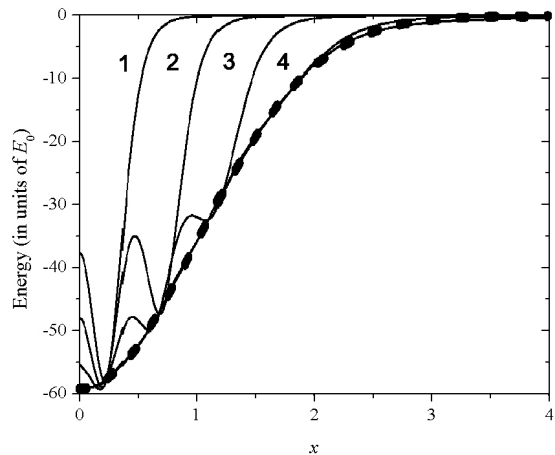, width=7cm, angle=270}
\caption{\small Comparison between different multisoliton
potentials. 1: two soliton, 2: four soliton, 3: six soliton, 4:
eight and ten soliton, bold dashed line represent the Gaussian
potential.}
\end{figure}
It is clearly seen from this figure the solitonic structure of the
curves for $N<8$ (curves 1, 2 and 3). The result for $N=8$ (curve
4) is not distinguishable of that for $N=10$ and practically
coincides with the given Gaussian potential curve. Moreover, for
$N=10$ the highest energy level is equal to zero with a high
precision. This indicates that this potential has only four
discrete levels which are not located near the ionization
threshold $E=0$ and possibly one level near this threshold.
Another remark worth making is that the long  distance behavior of
our potential is $\sim \exp (-A_0x)$ where $A_0=\sum a_j +\sum
b_j$ whereas the potential (\ref{pot}) tends to zero much faster,
as $\sim \exp (-A_1x^2)$.

\section{\label{results}Results}

\begin{figure}[t]
\centering \epsfig{file=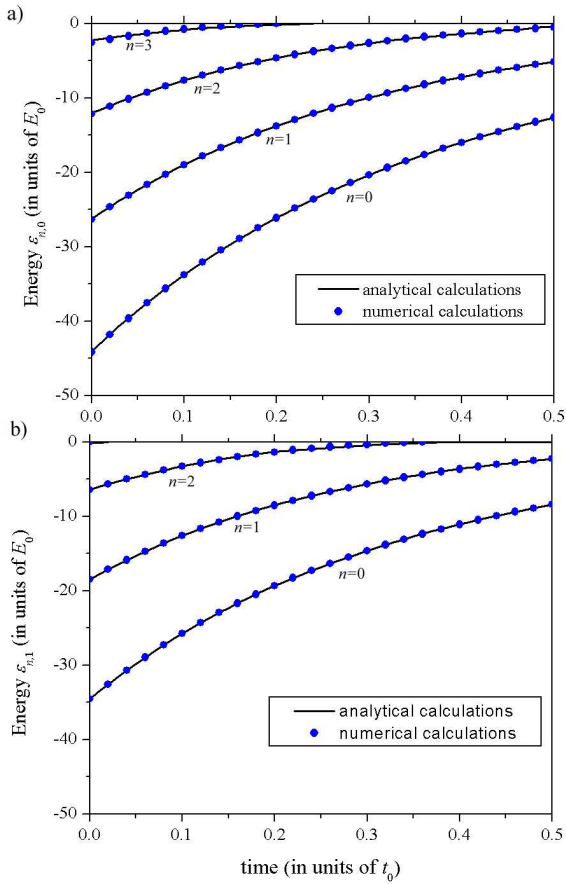, width=9cm} \caption{\small
(Color online) Energy spectra of electrons in NSPI QD with initial
half-width $d=1\mu$m and $B_{hf}(r=0,t=0)=2.65$T as a function of
time,
 $T_1/t_0=1$, $l=0$ (a) and $l=1$ (b). The solid lines are the energy levels
 obtained using the multisoliton approximate potential. The dots correspond to the energy
 levels, that were obtained as a numerical solution
 of the Shro\"odinger equation with the Gaussian potential.}\label{fig2}
\end{figure}

The time dependence of the confining hyperfine field given by Eq.
(\ref{B(r,t)}) determines the time-dependence of the electron
energy levels in the NSPI QD. There are two characteristic times
in the problem: the diffusion characteristic time $t_{0}$ and the
relaxation characteristic time $T_{1}$.
For $t \sim t_{0} \ll T_1$ we distinguish diffusive regime,
the times $t$ such that $t\ \sim t_0 \sim T_1 $ correspond to
  intermediate regime, and when  $t\sim T_{1} \ll t_{0}$
  we are in relaxation regime. Here $t$ is the observation time.

\begin{figure}[t]\label{fig3}
\centering \epsfig{file=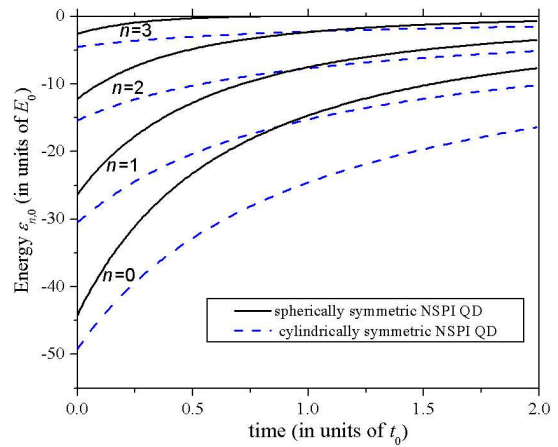, height=9cm, angle=270}
\caption{\small (Color online) Energy spectra of electrons in NSPI
QDs with spherical and cylindrical symmetry in the diffusive
regime, $T_1/t_0=100$. The parameters of calculations are as on
Fig. \ref{fig2}.}
\end{figure}

We found that the time-dependence of the energy levels in all
regimes are qualitatively similar for NSPI QD with spherically
symmetric confining potential: the number of energy levels, as
well as their depth, decreases with time. It is interesting to
note that the number of energy levels in cylindrically symmetrical
NSPI QD in the diffusive regime remains constant \cite{QD}, and
the number of transverse energy levels in NSPI QW in the diffusive
regime even increases with time \cite{QD}. This remarkable fact is
related to the symmetry of the confining potential: an increase of
the symmetry results in a faster spreading of the confining
potential due to the nuclear spin diffusion process.

In order to estimate the accuracy of our approximation, we
compared the results obtained by supersymmetric transformations
with numerical solutions of the Schr\"odinger equation with the
Gaussian potential. We have used the Shooting Method to solve the
radial Schr\"odinger equation, subjecting the solution to the
following boundary conditions:  $ R_{n,l} \left( r\rightarrow 0
\right)=r^{l}$ and
$R_{n,l} \left( r\rightarrow \infty \right)\to 0$
which correspond to a discrete spectrum eigenfunction.
It is convenient to define the function $P_{n,l}(r)$ as
$R_{n,l}(r)=r^lP_{n,l}(r)$ thus reducing the first of the above
conditions to
$(\partial P_{n,l}/\partial r)|_{r=0}=0$.

Figure 2 shows a representative result of our calculations. The
time-dependence of the energy levels was calculated for NSPI QD
based on GaAs with initial half-width $d=1\mu$m and
$B_{hf}(r=0,t=0)=2.65$T, corresponding to 50\% nuclear spin
polarization, in the intermediate regime, when  $T_1/t_0=1$. The
number of energy levels in the NSPI QD decreases with time. A
comparison between analytical and numerical results shows that the
method of supersymmetric transformations gives a good
approximation for almost all energies except for a tiny interval
near $E=0$. This is clearly seen from Fig. 2(a) for $n=3$ and Fig.
2(b) for $n=2$ energy levels. Such a difference occurs because of
different long distance asymptotics of the Gaussian and
multisoliton potentials, discussed at the end of the Section
\ref{method}.

Faster spreading of nuclear spin polarization in the case of 3D
diffusion than in the case of 2D diffusion results in
qualitatively different behaviour of the energy spectrum in NSPI
QDs with spherical and cylindrical symmetry in the diffusive
regime, as demonstrated in Figure 3 for $n=0$. It is seen from Fig.
3 (this was exactly proved in Ref. \cite{QD}) that the number of
energy levels in the cylindrically symmetrical NSPI QD in the
diffusive regime is constant in time. In the case of spherically
symmetric NSPI QD, the number of energy levels decreases in time.

\section{\label{conclusions}Conclusions}

We have investigated the electronic structure of a quantum dot
created through a spherically symmetric local nuclear spin
polarization within multisoliton potential approximation obtained
by means of Darboux transformations. In particular, the electron
energy spectrum of the NSPI QD having spherical symmetry was
calculated as a function of time. We found a specific feature of
the evolution of such NSPI QDs - the number of energy levels, as
well as their depth, decreases with time in all regimes. This
observation contrast with  previously studied cylindrically
symmetrical NSPI QD.

Moreover, it has been demonstrated that the approximation of a
Gaussian potential by a multisoliton potential is much more
efficient in comparison with the approximation by a parabolic
potential. The parabolic approximation is acceptable only for the
ground state, whereas multisoliton potentials give an excellent
agreement with the exact results for all energies except for a
tiny interval near $E=0$. Moreover, for an electron confined by a
multisoliton potential the wave function is given analytically. We
would like also to point out that the method presented here can be
applied for studying not only NSPI QDs but it also can be useful
for describing, for instance, traditional quantum dots.

\section*{Aknowledgments}

We acknowledge useful discussions with Professor V. Privman. This
research was supported in part by the National Security Agency and
Advanced Research and Development Activity under Army Research
Office contract DAAD-19-02-1-0035, and by the National Science
Foundation, grant DMR-0121146. B. F. S. acknowledges partial
support from President Grant of Russia 1743.2003.2, the Spanish
Ministerio de Education, Cultura y Deporte Grant SAB2000-0240, and
the Spanish MCYT and European FEDER grant BFM2002-03773. B. F. S.
is also grateful for the hospitality to the Center for Quantum
Device Technology of Clarkson University in the Autumn of 2003
where this work was started.

\end{document}